\documentclass[preprint,showpacs,amsmath,amssymo]{revtex4}
\usepackage[dvips]{epsfig}
\usepackage{bm}

\begin{document}

\title[c]{Vibrational and thermodynamic properties of high-pressure phases in LiYF$_{4}$}

\author{A. Sen}
\author{S. L. Chaplot}
\author{R. Mittal}
\affiliation{Solid State Physics Division, Bhabha Atomic Research Centre, Trombay, Mumbai 400 085, India}

\begin{abstract}
Possible variations in the dynamical behaviour of LiYF$_{4}$  due to its structural changes following several  pressure-induced phase  transitions are examined by making use of the complementary techniques of quasi-harmonic lattice dynamics and molecular dynamics simulation.  The phonon spectra in the entire Brillouin zone and the respective Gibbs free energies are calculated for the three high-pressure polymorphs of LiYF$_{4}$  that are stable at T = 0, with an aim to understand their relative stabilities as functions of pressure and temperature in terms of volume compression and vibrational entropy. Molecular dynamics simulations provide qualitative  impressions  about a temperature-driven second-order transformation and also of  kinetic effects in the subsequent pressure-driven first-order phase transition. In addition, the calculations predict anomalous thermal expansion at low temperature in phases I and IIa while irreversibilty of phase II $\rightarrow$ phase III transition on subsequent pressure release.
\end{abstract}

\maketitle

\section{INTRODUCTION}
Ternary halide  LiYF$_{4}$ (a laser host material of current interest\cite{saran, salaun, senprb1, man, senjpcm1, grez, cur, senprb2, sli, eran}) displays several structural phase transitions\cite{man, grez, senprb2} upon compression which turn out to be unique in comparison of isostructural CaWO$_{4}$. For example, while the latter has a scheelite ($I4_{1}/a$) to wolframite ($P2_{1}/c$) kind of high-pressure phase transformation, the former shows an intermediate fergusonite phase ($I2/a$, above room temperature; $P2_{1}/c$, in the low temperature region), associated with a soft phonon mode.\cite{senjpcm1} Recently, the post-fergusonite phase of LiYF$_{4}$ has been predicted as wolframitelike monoclinic one.\cite{senprb2, sli} Interestingly, the axial ratio in both the compounds evolves in a different way under the effect of pressure.\cite{eran} In  view  of the above,  we go for a comparative  analysis  of various dynamical properties that these phases (Fig.~\ref{fig1}) are likely to demonstrate. To reach this, we  carry  out  extensive  lattice  dynamical calculations  in  the  quasiharmonic  approximation (QHA) individually for the various low-temperature phases (e. g. I, IIa and III in Fig.~\ref{fig1}) of LiYF$_{4}$.

\vspace{0.2cm}
\parindent 0.8cm
With  the  advent  of  high-energy  synchrotron  X-rays  and sophisticated pressure cells, high-pressure lattice dynamics has gained considerable momentum in recent years  as  it helps  enable a direct comparison of experimental  findings, obtained either by the inelastic scattering processes or through some modern techniques of ultrafast X-ray diffraction.\cite{ultra} Our calculated  variation of phonon frequencies of  LiYF$_{4}$ in its scheelite phase as  a function   of   pressure  and  temperature (based   on   a transferable rigid ion model \cite{senprb1, senjpcm1}) has already been found to auger  well with the experimental  Raman  and infrared data.\cite{saran, salaun}

\vspace{0.2cm}
\parindent 0.8cm
It is well known that quasiharmonic lattice dynamics (QLD) has the ability to calculate \cite{born} vibrational free energies and other  derived  properties (e. g. entropy and specific heat) of a complex configuration even at very high pressures.  To achieve our goal, we make  use of  the  same  interatomic  potential as developed earlier \cite{senprb1} for lattice dynamical calculations. In an effort to strengthen the comparison of the  three  phases  of LiYF$_{4}$,  phonon  dispersions  and  phonon density of states have been examined. Gibbs free energies are then calculated for the three high-pressure phases LiYF$_{4}$  that are stable at T = 0, with an aim to understand their relative stabilities as functions of pressure and temperature in terms of volume compression and vibrational entropy. The present work aims at providing a possible pressure-temperature (P-T) phase diagram for LiYF$_{4}$ to help understand the underlying physics of phase transition in such compounds at the microscopic level. The second-order phase transitions and the kinetic effects in the first-order phase transitions are specifically investigated by molecular dynamics simulations.

\section{VIBRATIONAL PROPERTIES}
\subsection{Raman and infrared active modes}
While the scheelite phase($I4_{1}/a$, Z=4) of LiYF$_{4}$ has a total of 36  vibrational  degrees of freedom per primitive cell \cite{senprb1,senjpcm1,salaun}, the other two high-pressure  phases (IIa and III in Fig.~\ref{fig1}) possess  72  phonon  modes   at   each wavevector  due  to  doubling  of the primitive cell. Since both these high-pressure phases of LiYF$_{4}$ belong to the  same  space group($P2_{1}/c$,  Z=4),  the common irreducible representation of the phonons at the zone center is given by
\begin{center}
              $\Gamma$: 18$A_{g}$+18$A_{u}$+18$B_{g}$+18$B_{u}$,
\end{center}
of  which  A$_{g}$  and  B$_{g}$  modes  are Raman active while modes of ungerade symmetry (viz. A$_{u}$ and B$_{u}$) remain infrared active. It may be  noted  that the  degeneracy of the E$_{g}$ and E$_{u}$ modes which were present in phase I (scheelite) of LiYF$_{4}$ is lifted in  the  high-pressure  phases.  Table~\ref{tab1} yields  a  comparison of our calculated zone centre normal modes between phases IIa and III  at respectively  13 and 18 GPa. Phase IIa has a soft A$_{g}$ mode at 18 cm$^{-1}$ which hardens considerably to 92 cm$^{-1}$ in phase III. Most other modes harden slightly in going from phase IIa to phase III.
The vibrational frequencies for each of the phases are computed  by  diagonalization  of  the  respective  dynamical matrices using the software DISPR \cite{chap}, developed at Trombay.

\subsection{Phonon dispersion relations}
With a view to compare the phonon dispersion in various phases, we resort to the common high symmetry direction of $\Lambda$,  which  is  labeled as the $\bf \it c$ axis for phase I ($I4_{1}/a$) while it is the $\bf \it b$ axis for the other high-pressure(monoclinic) phases  of  similar space  group  symmetry (i. e. $P2_{1}/c$).  On undertaking group theoretical calculations, we  obtain  for the representations of all the normal modes in the three phases as follows:
\begin{center}
Phase I $\rightarrow$ $\Lambda$: 8$\Lambda_{1}$ + 8$\Lambda_{2}$ + 10$\Lambda_{3}$ ($\Lambda_{3}$ being doubly degenerate)\\
Phase IIa and Phase III $\rightarrow$ $\Lambda$: 36$\Lambda_{1}$ + 36$\Lambda_{2}$\\
\end{center}
Phonon dispersions of various polymorphs in LiYF$_{4}$ are displayed in Figure~\ref{fig2} along with the available inelastic neutron scattering data.\cite{salaun}
The experimental data\cite{salaun} for the phase I at zero pressure fit excellently with the lattice dynamical  calculations.\cite{senprb1}
Due to doubling of the unit cell in going from phase I to phase IIa, the  Brillouin zone in phase  IIa is halved and so there is an apparent folding back of the dispersion branches from zone boundary (of phase I) to zone center (of phase IIa). Consequently, there are several optic phonons at the zone center in phase IIa at low energies. Interestingly, we  come  across  a very low  energy zone-center  optic  mode  in  phase  IIa, indicating the onset of a possible high-pressure phase transition at  a pressure  close  to 13 GPa. This espouses our previous MD observations.\cite{senprb2}
The soft-alike phonon branch of phase IIa may be identified as  belonging to the longitudinal A$_{g}$ mode [Table~\ref{tab1}] as noted earlier. Further, a crossover between an  acoustic  phonon  mode and a low frequency optic mode is observed in phase IIa and also in phase III. We also notice that two acoustic phonon branches of $\Lambda_{2}$ symmetry in the phase III are too closely dispersed  to  remain  distinguishable.

\subsection{Phonon density of states}
To have an overview of the range and extent of various phonon modes in all the three phases, we resort to integrating over all the phonons with an energy resolution of 0.5 meV at each wave  vector  on  a  $16\times16\times16$ mesh  within the irreducible Brillouin zone. As Figure~\ref{fig3} suggests, the energy distribution extends by nearly 10 \% in switching from phase  I  to phase  IIa  while the change in the extent is negligible for the phase III transition. In order to check whether  it  is  due  to  simple  pressure effect,  we  compare  the respective frequency distribution for phase IIa and phase III at the same pressure of 13 GPa. One may notice that though the distribution patterns slightly differ, the  extent  remain  more  or less  the  same.  If one is further interested in looking for individual contributions  of  the  constituent  atoms  to  the   entire   frequency distribution,  partial  density  of states (PDOS) have to be estimated by atomic projections of the one-phonon eigenvectors.

\vspace{0.2cm}
\parindent 0.8cm
The labeling of the crystallographic axes viz. $a$, $b$, and $c$ varies as per convention adopted for specific space groups. In view of these differences and of the need to compare
the different polymorphs, we choose right-handed orthogonal  $x$, $y$ and $z$ axes as follows: $z$ along $\bf c$ and $x$ along $\bf a$ in phase I ($I4_{1}/a$); $z$ along $\bf b$ and $x$ along $\bf c$ in phases IIa and III ($P2_{1}/c$). Our  calculated  PDOS for various atoms in various phases are portrayed in Figure~\ref{fig4}.  $Y$ atoms are  found  to  contribute  largely  up  to  60  meV  while  $Li$ atoms make it (especially with polarization along  $x$ and  $y$) in the higher energy side of the density of states (DOS). However, $F$ atom contributions are  spread  over  the  entire energy  range  as if to replicate the total DOS to a considerable degree. Such variations in atomic contributions are partly due to the mass effect.

\section{Thermodynamic Properties}
\subsection{Heat capacity and Debye Temperature}
Following  the DOS calculations derived out of our simulated data for all  the  three  phases  of  LiYF$_{4}$,  we  compare  the respective  heat  capacities  at  constant pressure(C$_{P}$) as a function of
temperature. Figure~\ref{fig5}(a) demonstrates how    C$_{P}$ changes  for  different high-pressure  phases.  We  observe  that  although  initially  constant pressure heat capacity for phase III runs  lower,  it goes  up  above  the  room temperature and in the process even surpasses phase IIa. This may be because as we go on increasing temperature (T), the difference C$_{P}$(T) - C$_{V}$(T) [=$\alpha^2_{V}(T)BVT$]  (where C$_{V}$(T) is the constant volume heat capacity and $B$ is the isothermal bulk modulus) becomes significantly large, as inset of Figure~\ref{fig5}(a) suggests, due to the existence of much higher thermal expansion $\alpha_{V}(T)$ in phase III than that in phase IIa. A detailed discussion on $\alpha_{V}(T)$ follows in the next subsection.

\vspace{0.2cm}
\parindent 0.8cm
It is illustrative to compare the Debye temperature $\theta_D$(T) for the various phases [Figure~\ref{fig5}(b)] as derived from the calculated C$_{V}$(T).
We observe that in the very  low  temperature  region (say,  $<$ 20  K),  $\theta_D$ differs  a  lot from phase to phase (about 14 \% increase for I$\rightarrow$IIa and about 26 \% increase for IIa$\rightarrow$III) while at higher temperatures (say, 300 K),  it  comes
closer  for  phases  IIa and III as difference gets minimal (about 2 \% only).
\subsection{Gr\"uneisen parameter and thermal expansion}
Lattice  anharmonicity,  which  leads  to  a volume dependence of phonon frequencies($\omega_{i}$), is described by the mode Gr\"uneisen parameter \cite{grun}  given by
\begin {eqnarray}
\label{eq1}
\gamma_{i} = - {\partial ln~\omega_{i}\over\partial ln~V}
\end{eqnarray}
It  is  also the only kind of anharmonicity that can be taken care of within the framework of quasiharmonic  approximation.  Moreover, Gr\"uneisen parameter  is  an important quantity as  it  describes  the thermoelastic behaviour of materials at high pressures and temperatures. It  has  both  the  microscopic  and  macroscopic definition, the former
relating to the vibrational  frequencies  of  atoms  in  a  material (Eq.~\ref{eq1})  while the latter, to familiar thermodynamic properties such  as  heat  capacity  and  thermal  expansion. Unfortunately,   the
experimental  determination of $\gamma_{i}$, defined in either way, is often not so easy, since the microscopic definition requires a detailed knowledge  of the  phonon  dispersion  spectrum of a material, whereas the macroscopic
definition  calls  for  experimental   measurements   of   thermodynamic properties  at  high  pressures  and  temperatures. In this perspective, theoretical model calculation may be of great relief.

\vspace{0.2cm}
\parindent 0.8cm
Figure~\ref{fig6}  displays  our calculated  average $\gamma_{i}$ for various phonon energies and in various phases. It can be seen that below 10 meV, there is a significant number of  modes  with negative Gr\"uneisen parameter in phase I as well as phase IIa, while phase III modes have all positive $\gamma_{i}$.

\vspace{0.2cm}
\parindent 0.8cm
The effect of pressure on the volume coefficient of thermal expansion($\alpha_{V}$) can be studied through mode
Gr\"uneisen parameters($\gamma_{i}$) in the entire Brillouin zone.  In the quasiharmonic approximation, each phonon mode of energy E$_{i}$ contributes to the thermal expansion by way  of ($1 \over BV$)$\gamma_{i}$C$_{Vi}$, where $C_{Vi}$ denotes the specific heat (at constant volume) contribution of the $i^{th}$ mode and $V$ the lattice volume. This procedure suits well  to  the ambient  fluoroscheelite  system  because  explicit anharmonicity, which arises mainly out of thermal amplitudes is not very significant\cite{senjpcm1} compared to the implicit  effect  that  involves  phonon  frequency  change  with volume (as  obtained from $\gamma_{i}$).

\vspace{0.2cm}
\parindent 0.8cm
Figure~\ref{fig7}(a) demonstrates the variation of $\alpha_{V}$ as a function of temperature. It is also apparent that the scheelite as well  as  the initial high-pressure phase of LiYF$_{4}$ have negative thermal expansion in the low temperature limit (below 100 K). The reason may  be that  $\gamma_{i}$ has large negative values for phonons below 10 meV in these two phases. However, the third phase has no  such  anomaly, again because it has all positive gammas for phonons of all energies (Fig.~\ref{fig6}). Further, the  contributions  to  $\alpha_{V}$ from phonons of different energies (corresponding to various phases of LiYF$_{4}$) are displayed  in Figs.~\ref{fig7}(b), (c) and (d) respectively. We observe that at low temperatures (e.g. 20 K), contribution of modes upto 10 meV are quite significant to $\alpha_{V}$, but as temperature is increased (e. g. 300 K) higher energy modes get more populated and hence, contribute in a large way to the volume thermal expansion.

\subsection{Mean squared displacements and thermal anisotropy}
In  order  to  gain  some  insight  over how the phonons of various energies are polarized in various phases of LiYF$_{4}$, we plot in Figure~\ref{fig8} the partial contributions of these phonons to the  mean  squared  thermal
amplitude  of  the  constituent  atoms.  The mean squared displacement of atom $k$ along $\alpha$ direction is given by
\begin{equation}
U_{\alpha\alpha}(k,T) = \bigl\langle u^2_{k\alpha} \bigr\rangle_{T} = A{\hbar\over m_{k}} \intop\limits_{0}^{\infty} {\left(n+1/2\right)\over \omega} g_{k\alpha}(\omega)d\omega,
\label{eq3}
\end{equation}
where~$n$ = $\left[exp \left(\hbar\omega \over KT\right)-1\right]^{-1}$; $g_{k\alpha}(\omega)$  is  the partial density of states associated with an atom $k$ whose mass is m$_{k}$; $A$ is the normalization constant such that $\int g_{k\alpha}(\omega)d\omega$ = 1. As Figure~\ref{fig8}
suggests, the modes at very low energies involve equal displacements  of  all  the  atoms  that
correspond to the acoustic modes. Interestingly, between 2  and  9 meV, Y and F atoms have larger amplitudes than what relatively
light-weight  Li  atoms  possess.  It  may  be  noted  that   the   basic structure (scheelite)  of  LiYF$_{4}$ comprises a pair of strongly bonded LiF$_{4}$ tetrahedra and loosely bonded YF$_{8}$ polyhedra \cite{eran}. Larger amplitudes  of  F atoms  in  the  first  two  phases  mark  the  presence of librations of the  LiF$_{4}$  tetrahedra.

\vspace{0.2cm}
\parindent 0.8cm
Our calculated values of equivalent isotropic thermal parameters for $Li$, $Y$ and $F$ atoms in the ambient (i. e. P=0) scheelite phase  of LiYF$_{4}$ are found to be 17, 9 and 14 $\times$ 10$^{-2}{\rm \AA^{2}}$ as against the respective experimental\cite{garcia}  values of 20, 10 and 17 $\times$ 10$^{-2}{\rm \AA^{2}}$. A detailed comparison  of the anisotropic thermal parameters (U$_{\alpha\alpha}$) at different temperatures among  the different phases of LiYF$_{4}$ is further given in Table~\ref{tab2}. It  may be interesting to note that while $Y$ and $F$ atoms have comparable atomic displacements along the three directons in all the high-pressure phases of LiYF$_{4}$, $Li$ atoms, in contrast, show larger anisotropy along the $z$ direction in phase I, along the $y$ direction in phase II and again along the $z$ direction in phase III.

\subsection{Free energy and phase stability}
A  P-T phase  diagram  generally portrays various equilibrium phases at constant temperature (T) and  pressure (P)  with  the  lowest  Gibbs  free energy (G).   From   phonon   calculations,   the  temperature  dependent vibrational free energy at various  hydrostatic  pressures  for  various phases  of  LiYF$_{4}$  can  be  estimated in the quasiharmonic approximation. Thermodynamically, we may write\cite{hill}
\begin {equation}
G_{n}(P,T) = U_{n} - TS_{n} + P V_{n}
\label{eq4}
\end{equation}
where U$_{n}$, S$_{n}$~and V$_{n}$ refer respectively to the internal energy, the vibrational entropy and the lattice volume of the ${n}^{th}$  phase. This  free  energy  can be expressed as the sum total of configurational contributions(G$_{n}^{config}$) and vibrational contributions(G$_{n}^{vib}$). The former account  for mean atomic positions while the latter account for vibrations about the mean position.  Sometimes,  depending  on the material to be studied, contributions from magnetic effects(G$_{n}^{mag}$) and electronic effects(G$_{n}^{el}$) can  also be significant. However, since LiYF$_{4}$ is an insulator in all the phases being considered, we may use the Born-Oppenheimer approximation and therefore  we may not explicitly consider the electronic excitations. To  repeat  it  once  again,  the lattice  excitations  are  treated in this work within the quasiharmonic approximation  where  the  full  Hamiltonian  at  a  given   volume   is approximated  by  a  harmonic  expansion  about  the  equilibrium atomic
positions, though anharmonic effects are also included to a good  extent through  the  implicit volume dependence of the vibrational frequencies. Given this, the  vibrational  Gibbs  free  energy  (G$_{n}^{vib}$)  is  found  to  be
satisfyingly  accurate.

\vspace{0.2cm}
\parindent 0.8cm
To  include  vibrational  effects in the present phase diagram, we have calculated dynamical matrices separately  for  each  of  the  three phases (viz. I, IIa and III) at  pressure  intervals  of 2 GPa on a $4\times4\times4$ mesh in the irreducible Brillouin zone comprising 64 wave vectors. In order to rationalize the  behaviour  of  this  dynamical simulation,  we calculate \cite{asen} the enthalpy vs. pressure curves for all the three known structures of LiYF$_{4}$ and notice that the enthalpy changes (due to internal energy and volume) are predominant in pressure-driven transitions over free energy changes (due to vibrational energy and entropy). Perhaps  this  is  why  most  of  the
first-principles  phase   diagram   calculations of pressure-driven phase transitions \cite{duca}   do   not   include vibrational  entropies,  though configurational contributions are always taken into account. The other reason for this may  be  that  vibrational entropy  differences  between phases are assumed to be quite small as we come across in this particular case too.

\begin{figure}[h]
{\bf Figure Captions}
\caption{\label{fig1}Crystal structures of LiYF$_{4}$ belonging to (a) phase I, (b) phase IIa, (c) phase IIb and (d) phase III.}
\caption{\label{fig2}Comparison  of  phonon dispersion relations along the same
high  symmetry direction  of $\bf\Lambda$  in  the  three  phases  of  LiYF$_{4}$   at
respectively  0,  13  and  18  GPa. Lines refer to our calculations while  symbols represent the experimental data \cite{salaun}.}
\caption{\label{fig3}(a)Calculated phonon density of states, g(E), in the three
phases of LiYF$_{4}$ at respectively 4, 13 and  18  GPa;  (b)  comparison  of
g(E) in the two high-pressure phases at the same pressure of 13 GPa.}
\caption{\label{fig4}Calculated  partial  density of states for the constituent
atoms with contributions from polarizations along three orthogonal directions
in  the  three  phases  of LiYF$_{4}$ at respectively 4, 13 and 18 GPa. Since
phase I is of tetragonal symmetry,  the  contributions  along  $x$  and  $y$
directions are identical for this phase. The relationship between the orthogonal directions and the crystallographic axes is given in section IIC.}
\caption{\label{fig5}Calculated   (a)   heat   capacity (C$_{P}$)   and  (b)  Debye
temperature($\theta_{D}$) as a function of temperature for  the  three  phases  of
LiYF$_{4}$  at  respectively  4,  13  and  18  GPa. The inset depicts the difference of C$_{P}(T)$ and  C$_{V}(T)$
over the same temperature region.}
\caption{\label{fig6}Calculated  average mode  Gr\"uneisen parameter($\gamma_{i}$)as a function of
phonon energy for the three phases of LiYF$_{4}$ at respectively 4, 13 and 18 GPa.}
\caption{\label{fig7}Calculated  volume  thermal expansion coefficient ($\alpha_{V}$) as a
function of temperature for the three phases of LiYF$_{4}$ at respectively 4,
13 and 18 GPa. (b), (c) and (d) demonstrate  how  phonons  of  different
energies (in phase I, phase IIa and phase III respectively) contribute to $\alpha_{V}$
at 20 and 300 K.}
\caption{\label{fig8}Calculated thermal mean squared displacements  (on a semi-logarthmic scale) of
the constituent atoms for (a) phase I, (b) phase IIa and  (c)  phase  III
due  to  phonons  of  various  energies at 300 K. Insets in (a), (b) and (c) refer
to the same but on an expanded scale up to 10 meV for the respective phases of LiYF$_{4}$ at 4, 13 and 18  GPa. It may be noted that in phases IIa and III, there are four symmetrically different $F$ atoms, however, only the average contributions of all the $F$ atoms are shown here.}
\caption{\label{fig9}(a)Comparison  of  the  Gibbs  free energy per atom as a
function of pressure in all the three phases of LiYF$_{4}$, as obtained  from
the  lattice  dynamical  calculations at T = 300 K.  The  variation  in  Gibbs  energy
difference ($\Delta$G) is plotted in  (b)  with  pressure(P)  and  in  (c)  with
temperature(T)  giving a qualitative impression of the first order phase
transition that takes place in switching from phase IIa to a  dynamically
more  favourable phase III. Inset in (c) shows $\Delta$G vs. T up to 200 K on an
expanded scale.}
\caption{\label{fig10}Results of phase transitions as observed in MD simulations runs with increasing pressures at several constant temperatures (phase I $\rightarrow$ phase II(a/b) $\rightarrow$ phase III) and also with increasing temperatures at several constant pressures (phase IIa $\rightarrow$ phase IIb). The transitions among the phase I, IIa and IIb are found to be reversible while phase III retains on decompression to P=0. The dashed line indicates the phase boundary between phase II and phase III as determined by quasiharmonic free energy calculations.}
\end{figure}
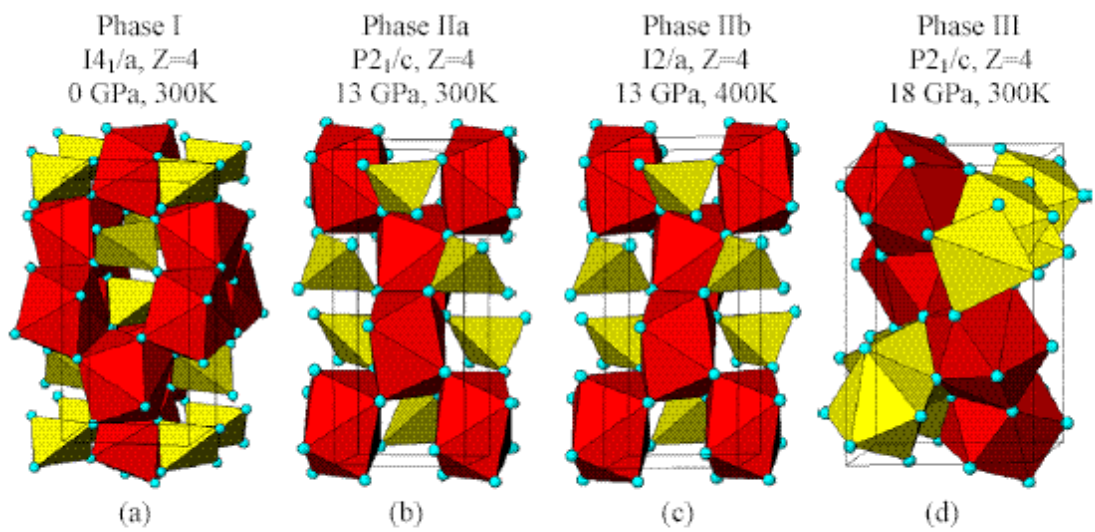
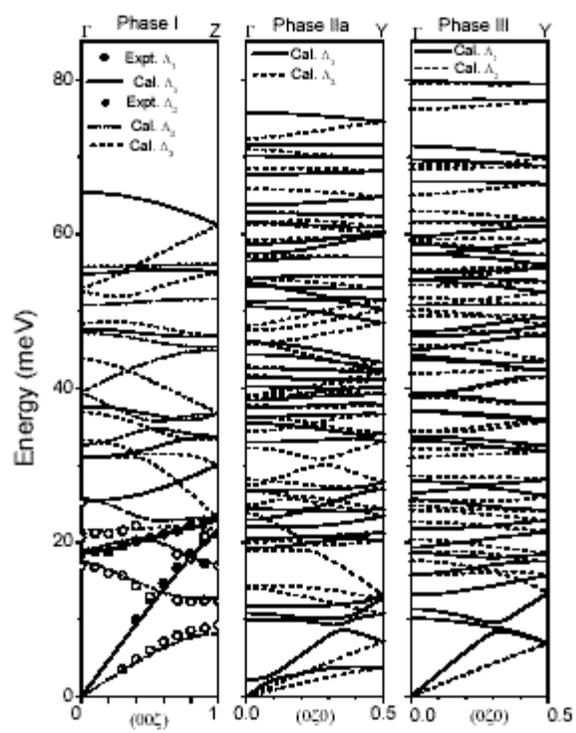
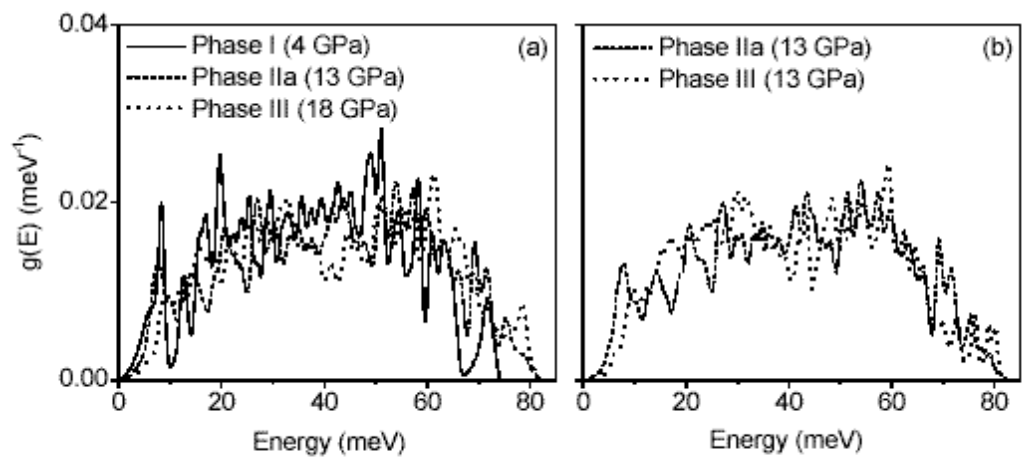
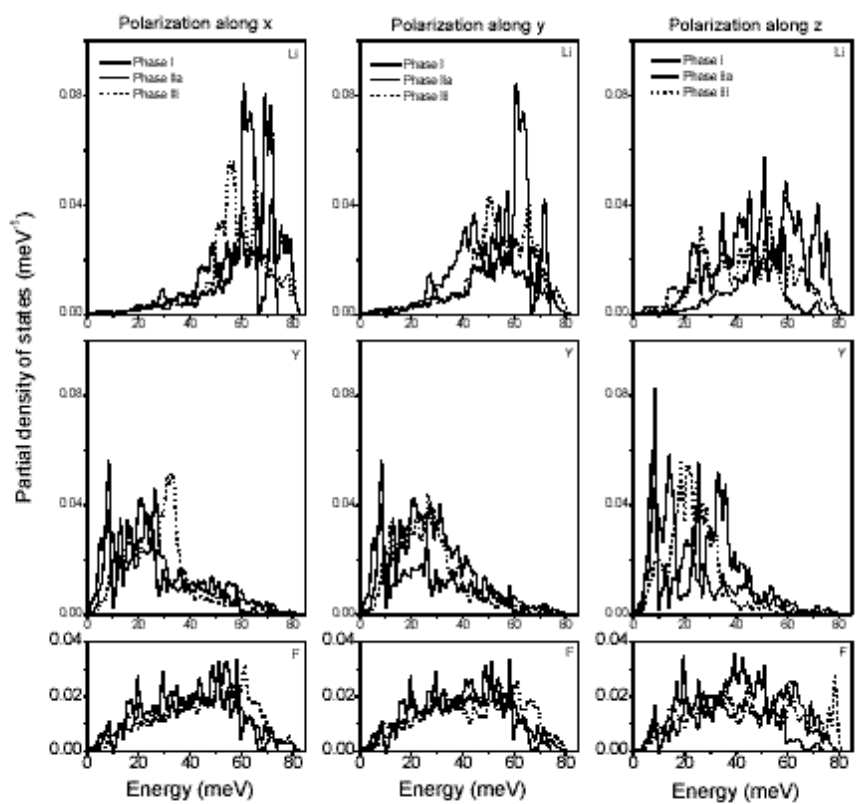
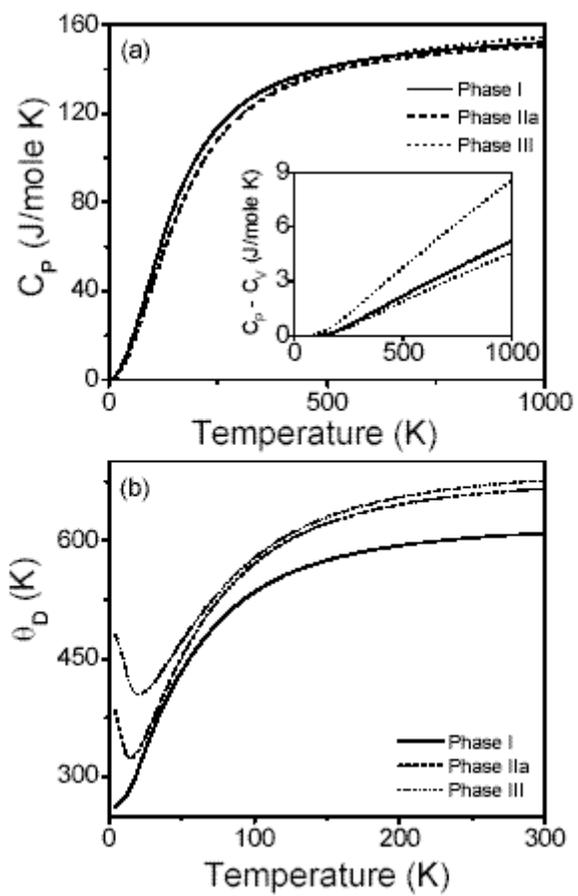
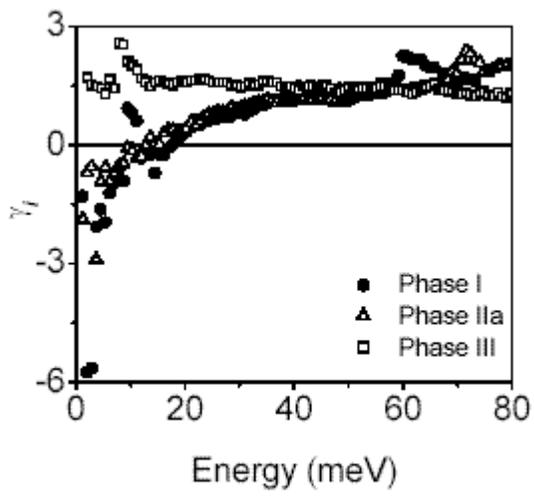
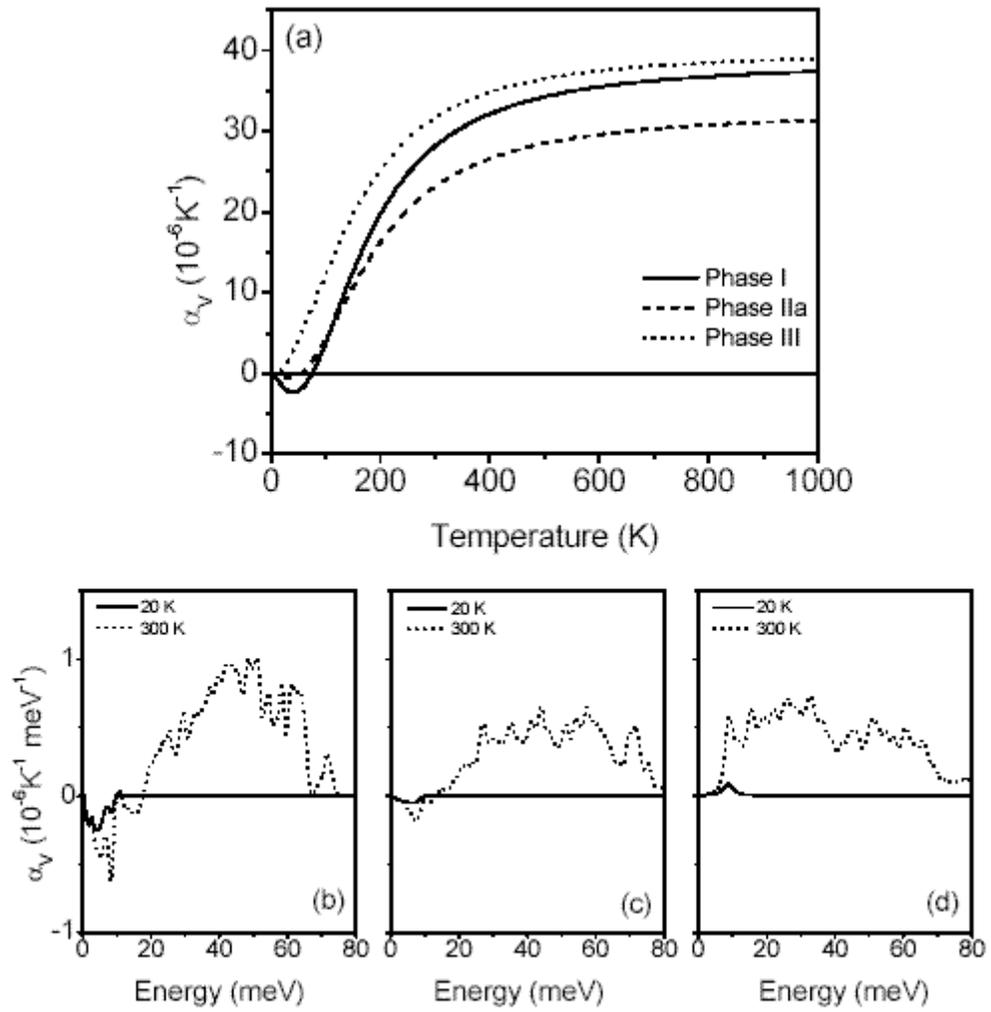
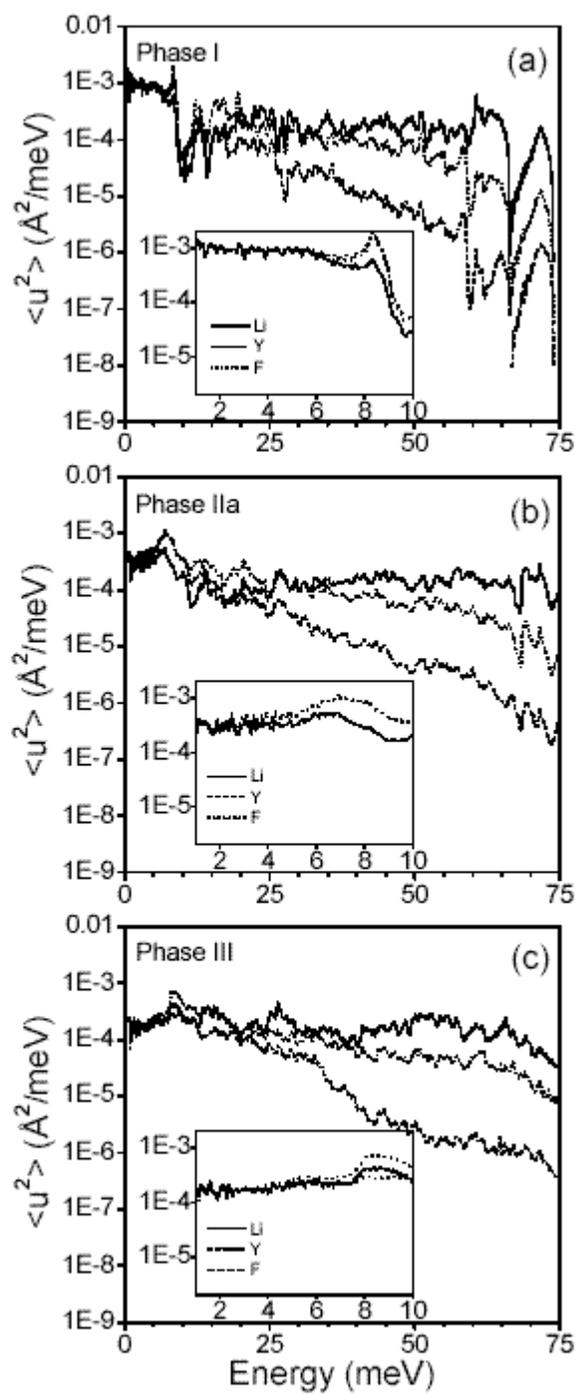
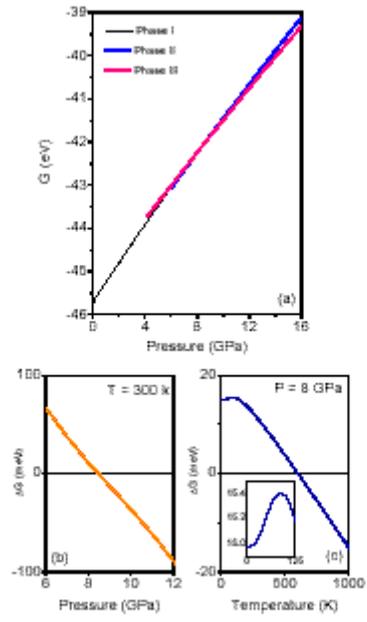
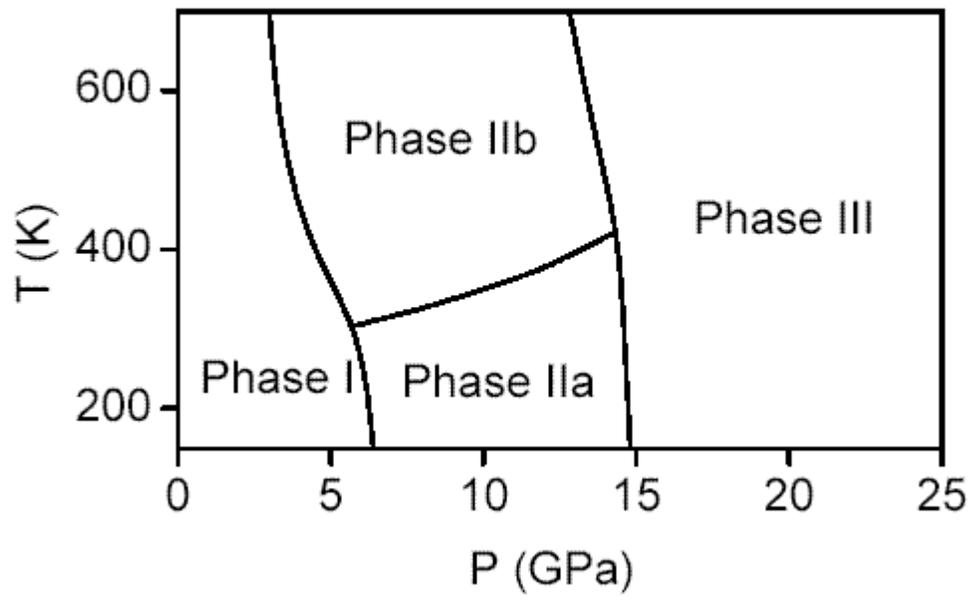

\vspace{0.2cm}
\parindent 0.8cm
The  stability  of a crystalline phase is largely determined by the minimization of the Gibbs free energy \cite{chapprb} and it would, hence, be  quite interesting  if we can compare the phase-wise free energy. The outcome is shown in Figure~\ref{fig9}. We note from Fig.~\ref{fig9}(a) that at 300 K the free  energy plot  of  the  phase I joins smoothly at 6 GPa to that of phase IIa, which is  consistent  with  the  nature  of  second  order  phase transition.  Beyond  8  GPa, phase III has a lower free energy indicating the greater stability of this phase and also the onset  of  a first  order  phase  transition  from  phase IIa to phase III. However, the transition pressure as obtained through the MD simulations\cite{senprb2} is higher than 8 GPa  due to hysteresis.  We observe  the greater stability of phase III at high pressures  arises  primarily  due  to  its  lower  volume,   while   the vibrational  energy remains also lower and the entropy becomes higher in phase III providing an additional stability. On the plot of  differences in  vibrational  Gibbs  free energy ($\Delta$G) as a function of pressure at 300 K, it is clearly seen that G for phase III is lower (above 8 GPa) than that for phase IIa. It is also observed (Fig.~\ref{fig9}(c)) that $\Delta$G between phase IIa  and  phase III structures decreases with temperature. Another interesting point to note from the plots of relative free energy vs. temperature (at 8 GPa) is that  in the very  low temperature region (say,  below  100  K), free energy difference shows an anomalous behaviour. This can be attributed to the fact that there is a greater density of low frequency modes in phase IIa than in phase  III due to the soft phonon modes in the former phase. We  get  further  support  in  this regard from Table~\ref{tab1}, where several vibrational modes have been compared between phase IIa and phase III.  At low  temperatures,  these  low  frequencies  are populated, thereby lowering the  free  energy.

\vspace{0.2cm}
\parindent 0.8cm
The above comparison of the Gibbs free energies in various phases easily provides the phase diagram involving the first order phase transitions. However, the second order transitions and also kinetic effects such as hysteresis in the first order transitions are better illustrated in molecular dynamics simulations. A comprehensive impression of the phase transitions in  LiYF$_{4}$ is given in Fig.~\ref{fig10} with increasing pressure and temperature (see caption of Fig.~\ref{fig10}). It strengthens our earlier observation\cite{senprb2} that phase II possesses two different space groups (viz. phase IIa: $P2_{1}/c$ ; phase IIb: $I2/a$) existing in two different ranges of temperaure. The low temperature phase IIa involves small displacements of some of the $F$ atoms from their ideal positions of body-centered symmetry in phase IIb.  The transition IIa $\rightarrow$ III occurs at a much higher pressure of 14 GPa than the equilibrium phase boundary at 8 GPa due to hysteresis at the time scale of the simulation and is found irreversible on release of pressure.

\section{Conclusions}
We have been able to demonstrate how a conjugate $\it {lattice}$ $\it {dynamics}$ - $\it {molecular}$ $\it {dynamics}$ study can eventually lead one to calculate the vibrational free energy and other essential thermodynamic  functions of LiYF$_{4}$ in all its available high-pressure phases from the mere knowledge of an interatomic pair potential. It also sheds light on the useful understanding of some important physical properties and phenomena associated  with  the phase equilibria. Finally, our  calculated results  have the potential ingredients to help  analyze  the  experimental inelastic scattering data,  if  high-pressure (along with low temperature) phonon measurements are carried out in future, which could be affirmative as well as interesting too.

\begin{acknowledgments}
A.  S.  would  like  to  express his deep sense of gratitude to the Council  of  Scientific  and  Industrial  Research  (CSIR, New Delhi),  India,  for rendering necessary financial assistance throughout the work and acknowledge  as  well the continuous encouragement and care taken by Dr.   M.   Ramanadham and Dr. V. C. Sahni.
\end{acknowledgments}

\newpage
\begin{table*}
\caption{\label{tab1}Comparison of calculated zone-center phonon frequencies between
phase IIa  (P= 13 GPa) and phase III (P=18 GPa) of LiYF$_{4}$ at T=0 K.}
\begin{ruledtabular}
\begin{tabular}{cccccccc}
\multicolumn{8}{c}{Long wavelength optical phonon modes(cm$^{-1}$)}\\
\multicolumn{2}{c}{A$_{g}$}&\multicolumn{2}{c}{B$_{g}$}&\multicolumn{2}{c}{A$_{u}$}&\multicolumn{2}{c}{B$_{u}$}\\
Phase IIa&Phase III&Phase IIa&Phase III&Phase IIa&Phase III&Phase IIa&Phase III\\
\hline
                18&92&115&127&0&0&0&0\\
            95&107&162&151&80&82&0&0\\
    155&151&193&165&87&128&115&142\\
166&175&217&200&185&156&152&183\\
178&214&227&250&196&206&199&195\\
215&225&260&277&266&268&220&205\\
286&271&293&314&293&298&277&230\\
303&313&316&316&310&299&304&259\\
324&317&336&398&343&352&310&278\\
334&363&369&404&358&356&315&337\\
371&377&384&417&408&382&345&368\\
413&410&438&431&434&436&370&383\\
429&436&461&447&455&463&388&410\\
435&443&496&476&463&473&437&435\\
473&463&507&478&495&497&477&495\\
506&481&531&509&514&539&494&526\\
546&562&577&614&578&623&553&552\\
565&575&584&640&610&643&572&556\\
\end{tabular}
\end{ruledtabular}
\end{table*}
\begin{table*}
\caption{\label{tab2}Calculated anisotropic thermal parameters (in units of 10$^{-4}{\rm \AA^{2}}$) for  the constituent atoms at various temperatures of 20 and 300 K associated with the three high-pressure phases of LiYF$_{4}$. It may be noted that in the scheelite phase ($I4_{1}/a$, Z=4), all the $F$ atoms are symmetrically equivalent.  See text for the labeling of $x$, $y$, $z$ directions.}
\begin{ruledtabular}
\begin{tabular}{ccccccccccc}
Species&Temperature&\multicolumn{3}{c}{Phase I (P=4 GPa)}&\multicolumn{3}{c}{Phase IIa (P=13 GPa)}&\multicolumn{3}{c}{Phase III (P=18 GPa)}\\
&(K)&U$_{xx}$&U$_{yy}$&U$_{zz}$&U$_{xx}$&U$_{yy}$&U$_{zz}$&U$_{xx}$&U$_{yy}$&U$_{zz}$\\
\hline
Li&20&63&63&88&54&72&60&58&64&85\\
&300&137&137&224&98&144&109&95&114&189\\
Y&20&17&17&14&13&12&17&11&12&13\\
&300&107&107&71&63&52&94&39&42&51\\
F(1)&20&41&41&40&36&36&36&29&28&42\\
&300&154&154&124&108&103&112&70&130&106\\
F(2)&20&&&&36&38&42&33&30&34\\
&300&&&&112&113&162&88&88&106\\
F(3)&20&&&&36&36&38&31&36&31\\
&300&&&&110&108&121&78&78&106\\
F(4)&20&&&&35&36&35&34&45&36\\
&300&&&&103&100&113&93&127&106\\
\end{tabular}
\end{ruledtabular}
\end{table*}
\newpage
\end{document}